\documentclass[8.5pt,twoside,twocolumn]{article}
\usepackage[super,sort&compress,comma]{natbib} 
\usepackage{mhchem}
\usepackage{bm}
\usepackage{amsmath}
\usepackage{amssymb} 
\usepackage{amsthm}
\usepackage{times,mathptmx}
\usepackage{sectsty}
\usepackage{balance} 
\usepackage{color}

\usepackage{graphicx} 
\usepackage{lastpage}
\usepackage[format=plain,justification=raggedright,singlelinecheck=false,font=small,labelfont=bf,labelsep=space]{caption} 
\usepackage{fancyhdr}
\newcommand{\red}[1]{\textcolor{black}{#1}}
\newcommand{\blue}[1]{\textcolor{black}{#1}}
\newcommand{\green}[1]{\textcolor{black}{#1}}
\newcommand{\magenta}[1]{\textcolor{black}{#1}}
\begin{document}

\thispagestyle{plain}
\fancypagestyle{plain}{
\fancyhead[L]{\includegraphics[height=8pt]{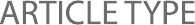}}
\fancyhead[C]{\hspace{-1cm}\includegraphics[height=20pt]{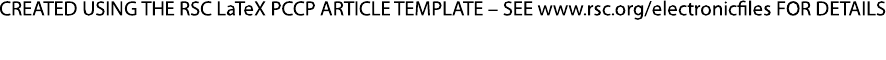}}
\fancyhead[R]{\includegraphics[height=10pt]{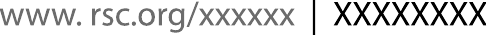}\vspace{-0.2cm}}
\renewcommand{\headrulewidth}{1pt}}
\renewcommand{\thefootnote}{\fnsymbol{footnote}}
\renewcommand\footnoterule{\vspace*{1pt}%
\hrule width 3.4in height 0.4pt \vspace*{5pt}} 
\setcounter{secnumdepth}{5}

\makeatletter 
\def\subsubsection{\@startsection{subsubsection}{3}{10pt}{-1.25ex plus -1ex minus -.1ex}{0ex plus 0ex}{\normalsize\bf}} 
\def\paragraph{\@startsection{paragraph}{4}{10pt}{-1.25ex plus -1ex minus -.1ex}{0ex plus 0ex}{\normalsize\textit}} 
\renewcommand\@biblabel[1]{#1}            
\renewcommand\@makefntext[1]%
{\noindent\makebox[0pt][r]{\@thefnmark\,}#1}
\makeatother 
\renewcommand{\figurename}{\small{Fig.}~}
\sectionfont{\large}
\subsectionfont{\normalsize} 

\fancyfoot{}
\fancyfoot[LO,RE]{\vspace{-7pt}\includegraphics[height=9pt]{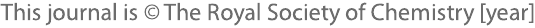}}
\fancyfoot[CO]{\vspace{-7.2pt}\hspace{12.2cm}\includegraphics{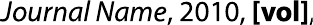}}
\fancyfoot[CE]{\vspace{-7.5pt}\hspace{-13.5cm}\includegraphics{RF}}
\fancyfoot[RO]{\footnotesize{\sffamily{1--\pageref{LastPage} ~\textbar  \hspace{2pt}\thepage}}}
\fancyfoot[LE]{\footnotesize{\sffamily{\thepage~\textbar\hspace{3.45cm} 1--\pageref{LastPage}}}}
\fancyhead{}
\renewcommand{\headrulewidth}{1pt} 
\renewcommand{\footrulewidth}{1pt}
\setlength{\arrayrulewidth}{1pt}
\setlength{\columnsep}{6.5mm}
\setlength\bibsep{1pt}

\twocolumn[
  \begin{@twocolumnfalse}
\noindent\LARGE{\textbf{Dynamics of Self-Threading Ring Polymers in a Gel}}
\vspace{0.6cm}

\noindent\large{\textbf{Davide Michieletto\textit{$^{a}$}, Davide Marenduzzo\textit{$^{b}$},
Enzo Orlandini\textit{$^{c}$}, 
Gareth P. Alexander{$^{a}$},
Matthew S. Turner{$^{\ast}$}{$^{a}$}}}\vspace{0.5cm}

\noindent\textit{\small{\textbf{Received Xth XXXXXXXXXX 20XX, Accepted Xth XXXXXXXXX 20XX\newline
First published on the web Xth XXXXXXXXXX 200X}}}

\noindent \textbf{\small{DOI: 10.1039/b000000x}}
\vspace{0.6cm}

\noindent \normalsize{We study of the dynamics of ring polymers confined to diffuse in a background gel at low concentrations. We do this in order to probe the inter-play between topology and dynamics in ring polymers. We develop an algorithm that takes into account the possibility that the rings hinder their own motion by passing through themselves, \emph{i.e.} ``self-threading''. Our results suggest that the number of self-threadings scales extensively with the length of the rings and that this is substantially independent of the details of the model. The slowing down of the rings' dynamics is found to be related to the fraction of segments that can contribute to the motion. Our results give a novel perspective on the motion of ring polymers in gel, for which a complete theory is still lacking, and may help us to understand the irreversible trapping of ring polymers in gel electrophoresis experiments.    }
\vspace{0.5cm}
 \end{@twocolumnfalse}
  ]

\section{Introduction}

\footnotetext{\textit{$^{a}$ Department of Physics and Centre for Complexity Science, University of Warwick, Coventry CV4 7AL, United Kingdom.}}
\footnotetext{\textit{$^{b}$ School of Physics and Astronomy, University of Edinburgh, Mayfield Road, Edinburgh EH9 3JZ, Scotland, United Kingdom.}}
\footnotetext{\textit{$^{c}$ Dipartimento di Fisica e Astronomia, Sezione INFN, Universit\`a di Padova, Via Marzolo 8, 35131 Padova, Italy.}}

The dynamics of large polymer molecules diffusing in a gel plays a central role in polymer physics and biology. The motion of linear and branched polymers in solution have been thoroughly studied in the past (see \cite{Gennes1979a, Doi1988} and references therein). On the other hand, there has been much less progress in understanding the motion of closed (ring) polymers in a gel \cite{Klein1986, Cates1986, Rubinstein1986, Obukhov1994}. The dynamics of ring polymers, because of the lack of ends, differs markedly from those of their linear cousins, involving fundamentally different modes of stress relaxation \cite{Kapnistos2008} and diffusion \cite{Robertson2006, Robertson2007, Halverson2011a, Michieletto2014}. A proper understanding of ring polymers, and the associated non-local topological constraints that they must satisfy, remains one of the major unresolved challenges in polymers physics. These properties have particular relevance to DNA, which can occur in circular form in Nature, \emph{e.g.} as bacterial plasmids, and its characterisation by gel electrophoresis \cite{Deutsch1988, Alon1997, Trigueros2001,Rosa2008}. In order to be able to give a satisfactory interpretation of these experiments, a deeper understanding of the mechanisms driving the diffusion of the ring polymers at equilibrium is required.

\begin{figure*}[t]
\centering
\includegraphics[scale=0.26]{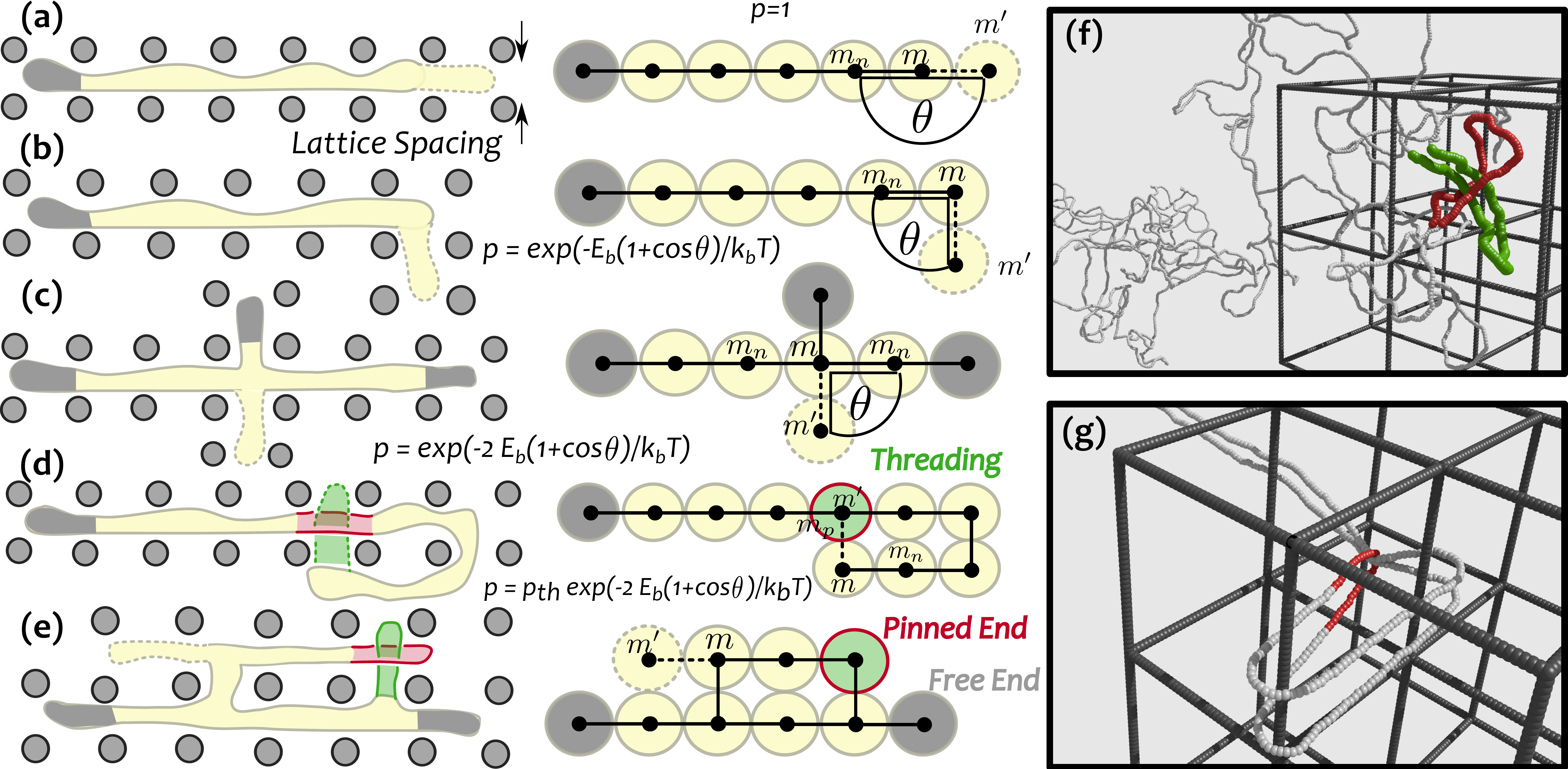}

\caption{(Colour online). Cartoon picturing the dynamics of the rings. The rings in the gel (left) are represented by lattice animals moving on the dual of the gel lattice (centre). \textbf{(a) (b)} and \textbf{(c)} Represent the moves allowed in a gel: translation, bending and branching, each by retraction of a terminal segment (shaded in grey). See text for more details. \textbf{(d)} With probability $p_{th} p_{m \rightarrow m^\prime}^{branch/bend}$ a bead can move onto a site that is already occupied to become an effective \emph{pin} (green) for the occupant bead $m_{p}$ (red). \magenta{Computationally, we implement this by stacking two beads on top of each other, on the same site (green bead on top of red one)}. \textbf{(e)} This is the case in which the threaded segment coincides with an end, in this case such end becomes ``pinned''. \magenta{In this configuration, the red segment/bead cannot be retracted, while the green one can contribute to the motion by annihilating with an anti-kink (see text). In this case $m^\prime$ can extend only if one of the grey beads or the green one is removed. In the case the red one is attempted to be retracted, the move would be rejected.} \textbf{(f)} and \textbf{(g)} show two snapshots of Molecular Dynamics simulation showing a self-threading and a self-trapping configurations, respectively. The colors highlight different segments of the chain. In (f) an ending segment of a branch (green) threads through another ending segment of another branch (red). This case is analogous to case (e) in the left panel. Fig. (g) is obtained after a strong electric field is applied to a self-threading configuration. One can imagine applying a field directed upward to the configuration in (d). The green segment may elongate, while the grey (free) end slides backward until it coincides with the red segment. If this occurs, the configuration is ``trapped'', like the one pictured in (g). The gel structure in (f) and (g) is sketched only partially and thinned for simplicity. \magenta{(see S.I. for details)}}
\label{fig:Hop}
\end{figure*}

The configuration of sufficiently long unknotted self-avoiding ring polymer unlinked from the gel is a double-folded self-similar branched tree (see left side in Fig.~\ref{fig:Hop}), also called a lattice animal \cite{Lubensky1979,Soteros1988,Obukhov1994,Whittington1982}. It is well-known that the gyration radius of a self-avoiding lattice animal grows as $M^{\nu}$, with $\nu = 1/2$ as opposed to $\nu \simeq 0.588$ for the case of a linear swollen coil in three dimensions \cite{Parisi1981}. Because of this, the self-density $\rho_s \equiv M/\langle R_g^2 \rangle^{3/2}$ of a large ring polymer in gel is much higher than that of either rings in good solvent, or linear polymers. Because of the compactness of the configurations, the contact probability between different segments belonging to the same chain, is higher. This implies that ring polymers in gel are more likely to hinder their own motion by interacting with themselves, as previously speculated~\cite{Klein1986,Obukhov1994}. In the following, we will be interested in configurations of a ring in gel in which a double-folded segment opens up and is threaded by another double-folded segment of the same chain (``self-threading'') (see Fig.~\ref{fig:Hop}(f)). In particular, when a ring polymer is forced to move inside a gel, self-threading can hinder polymer diffusion. Imagine a ring that winds around a strand of the gel and passes through itself, as in Fig. \ref{fig:Hop}(f). In this case the threading segment (green) behaves as a temporary ``pin'' for the threaded one (red), because of the uncrossability constraint. In order for the latter to freely diffuse, the former has to be removed. In the limit of large rings, one can think of a growing number of penetrations which can assume a hierarchical structure (imagine a segment of the polymer that threads through another that then is itself threaded, etc.). These would have to be undone in order to re-establish free diffusion, consequently increasing the polymer relaxation time. Self-threading is also a candidate for describing the low electrophoretic mobility (at low fields) and irreversible trapping (at high fields) of long ring molecules \cite{Viovy1992,Akerman1998,Viovy2000}. Imagine applying an electric field to the configuration in Fig.~\ref{fig:Hop}(f), in the direction parallel to the green segment. The ring can end up ``tightening'' itself around the obstacle, and become irreversibly self-trapped (see Fig.~\ref{fig:Hop}(g)), in a process that resembles the tightening of a knot \cite{Pieranski2001}. Threading has been recently acknowledged to be a non-negligible topological constraint in the case of ring polymers in dense solutions \cite{Lo2013,Bernabei2013,Michieletto2014}, which can affect both static and dynamic properties of the system. 

Here, we study how threadings of the chains through themselves can affect their own motion in dilute conditions. We study ring polymers diffusing in a gel by coarse graining the double-folded configurations to a network of beads located in the  cells of the gel (see Fig.~\ref{fig:Hop}). This coarse graining procedure maps the problem of simulating ring polymers diffusing in a gel to that of annealed branched polymers diffusing on a lattice, or lattice animals\cite{Lubensky1979}, under a specific set of rules which preserve the rings' topology (see Fig.~\ref{fig:Hop} and next section). The novel aspect of this work is in how we deal with the dynamics. This is simulated using an equivalent model to the kink-gas diffusion introduced by de Gennes \cite{DeGennes1971,Cates1986,Obukhov1994}, suitably modified to correctly take into account the slowing down due to the chains self-threadings. Our results suggest that in the limit of large rings, self-threadings increase extensively with the rings length and the dynamics is consequently slowed down.

\section{Algorithm and Computational Details}
We implement the coarse-grained model by means of a Lattice Kinetic Monte-Carlo simulation of \green{isolated} lattice animals formed by $M$ beads diffusing on a lattice. Along with the coarse-grained model, we perform a Molecular Dynamics simulation of a single ring polymer consisting of $M=5120$ beads (see S.I. for details), immersed in a gel which has lattice spacing equal to the ring's Kuhn length (see Fig.~\ref{fig:Hop}(f) and (g)). The latter is only intended to investigate static configurations of a large polymer in gel with fixed length. Static and dynamic properties for rings with different length are then left to be studied by means of the coarse-grained Monte Carlo model. 

In order to simplify our model we make some assumptions: Firstly, by replacing a double-folded segment of the polymers with a single bead filling a unit cell, we are implicitly assuming that the lattice spacing $l$ of the gel is comparable to the rings' Kuhn length $l_k$. The polymers are, therefore, flexible on the length scale of the gel pores. By making this choice, we also assume that the gel is tighter than a typical 6 \% agarose gel by a factor of 2~\cite{Pernodet1997}. Capillary or polyacrylamide gels electrophoresis offer interesting scenarios in which the Kuhn length of the samples is comparable with the pore size~\cite{Viovy2000}. Also, hydro-gels made from DNA strands and joints~\cite{Um2006,Park2009,Lee2012} have highly tunable properties and a very high tensile modulus that could produce materials with pore sizes comparable to the Kuhn length of the polymers analysed. We simplify further our model by assuming that the gel fibers are rigid.

\green{We study systems of $N=10$ non-interacting ring polymers of length $M=32$, $64$, $128$, $256$, $512$ beads. The rings do not interact with each-other and therefore behave as if they were isolated. We perform at least 3 realisation per simulation, meaning that we average at least over $3N$ rings}. The rings are prepared unlinked from the gel, and therefore they must assume a double-folded configuration, \emph{i.e.} every unit cell of the gel has both an out-going and in-going polymeric strand (see Fig.~\ref{fig:Hop}). We coarse-grain the polymers and represent both the out-going and in-going strands with one bead, which has the size of a Kuhn segment, and that spans the entire unit cell. In this way a ring with $2M$ segments is modelled via a collection of $M$ beads. The rings are treated as lattice animals diffusing on a cubic lattice which is shifted by $(\frac{l}{2}, \frac{l}{2}, \frac{l}{2})$ with respect to the gel lattice \cite{Weber2006a}, which is also modelled as a perfect cubic lattice. In other words, we model the rings by tracking the backbone and the branches of the lattice animal shapes they take (see Fig.~\ref{fig:Hop}). Our algorithm penalises the creation of new branches and the bending of the terminals by introducing two energies, $E_{branch} = 2 E_b (1 + \cos{\theta})$ and $E_{bend} = E_b(1+\cos{\theta})$ respectively, where $\theta$ is the angle formed by consecutive pairs of beads and $E_b = k_b T$. The motivation for this is twofold: (1) pure translation does not involve any change in energy (Fig.~\ref{fig:Hop}(a)) and (2) branching involves a creation of a new double folded terminal segment which, in terms of angles, contains two $90^\circ$ angles with the neighbours ($m_n$ in Fig. \ref{fig:Hop}) and one (newly formed) $180^\circ$ angle (being a terminal segment). Therefore, we approximate the energy penalty as twice the energy taken to bend an existing terminal segment (see Fig.\ref{fig:Hop}(b) and (c)). 

Our algorithm is the following: first we pick one bead ($m$) randomly. Second, we attempt a move to occupy a neighbouring site ($m^\prime$): 
\begin{itemize}
\item If $m^\prime$ is free, then the move is tested by means of the Metropolis algorithm, where the probability to be accepted is given by 
\begin{equation}
p_{m \rightarrow m^\prime}^{branch/bend} = \exp\{-E_{branch/bend}(\theta)/k_b T\}    \notag
\end{equation}
\item If the site $m^\prime$ is occupied by a non-neighbouring bead of $m$ then the move is accepted with probability:
\begin{equation}
p_{m \rightarrow m^\prime}^{branch/bend} = p_{th} \exp\{-E_{branch/bend}(\theta)/k_b T\}  \notag
\end{equation}
where $p_{th}$ is a free parameter in our model and represents the probability of self-threading.  
\end{itemize}
In order to correctly reproduce the hindering of the motion when the terminal segments are threaded, and hence not free to diffuse and contribute to the motion we adopt the following strategy: Once that the move has been ``energetically'' accepted, we place a new virtual segment at $m^\prime$ (represented as a dotted segment/circle in Fig.~\ref{fig:Hop}) and simulate the contour diffusion of an ``anti-kink'' (or ``hole'') that starts from $m$ and can annihilate only with one of the terminal beads of the lattice animal (represented as segments/circles shaded in grey in Fig.~\ref{fig:Hop}) or the newly formed segment $m^\prime$. For instance, in Fig. \ref{fig:Hop}(a), once that $m^\prime$ is created, a random walk starts from $m$ and can hit either $m^\prime$ (which happens most of the times), or the grey bead at the other end of the chain. In the first case, the chain does not move ($m^\prime$ is created and removed). In the second case, the chain steps to the right by one site ($m^\prime$ is created and the grey bead removed).    
Instead of having ``kinks'' (or segments with stored length) that accumulate along the contour and diffuse until they stop by extending a new segment, we first extend a new segment and then look for a terminal segment which can be retracted in order to accommodate the newly formed protrusion (and conserve the total mass). Since only one kink per time is allowed to travel along the lattice animal, \emph{i.e.} kinks do not interact, our method is completely equivalent to the kink-gas diffusion introduced by de Gennes \cite{DeGennes1971,Cates1986,Obukhov1994}, with the difference that our model can take into account long-ranged correlations which are essential in the motion of polymers with a closed topology. In de Gennes' picture, the polymers moved by accumulating mass, or length defects (kinks) along the contour, and by randomly spreading the excess of mass towards the terminal segments, which can extend. Here, we allow for temporary extensions of the terminal segments by creating a pair mass-hole (``kink'' - ``anti-kink''). The former immediately settles at the end of the segment ($m^\prime$), the latter starts a random walk along the chain and stops when either an end, or $m^\prime$ itself is hit (see next section for details). 
\magenta{It is also worth noting that we implicitly assume that the probability of unthreading is $1$ and that any energy barrier for unthreading is small, noting that the barrier for threading is likely to be much larger than the barrier for unthreading\footnote{We can actually generalise our interpretation by noting that the average number of penetrations will be determined by the ratio of probabilities (difference in energy barriers) between threading and unthreading, as usual. An alternative interpretation of $p_{th}$ would therefore be the ratio of these rates, so that each value would correspond to the thermodynamically correct density of penetrations. The actual kinetic rate for unthreading, when the end of the duplex ring has already diffused to the site of the threading, would still neglect the effect of a relatively small energy barrier. However, we believe this to be a tolerable simplification within a general philosophy that involves aggressive simplification, particularly given that the unthreading dynamics are likely dominated by the rate of diffusion of the penetrating portion of the duplex ring, which can be much larger than a single unit. A small correction to the kinetic rate constant for the final unthreading step would then yield an even smaller correction to the overall result for the (un)threading dynamics. We therefore neglect it entirely for simplicity.}.}


\begin{figure}[t]
\includegraphics[scale=0.3]{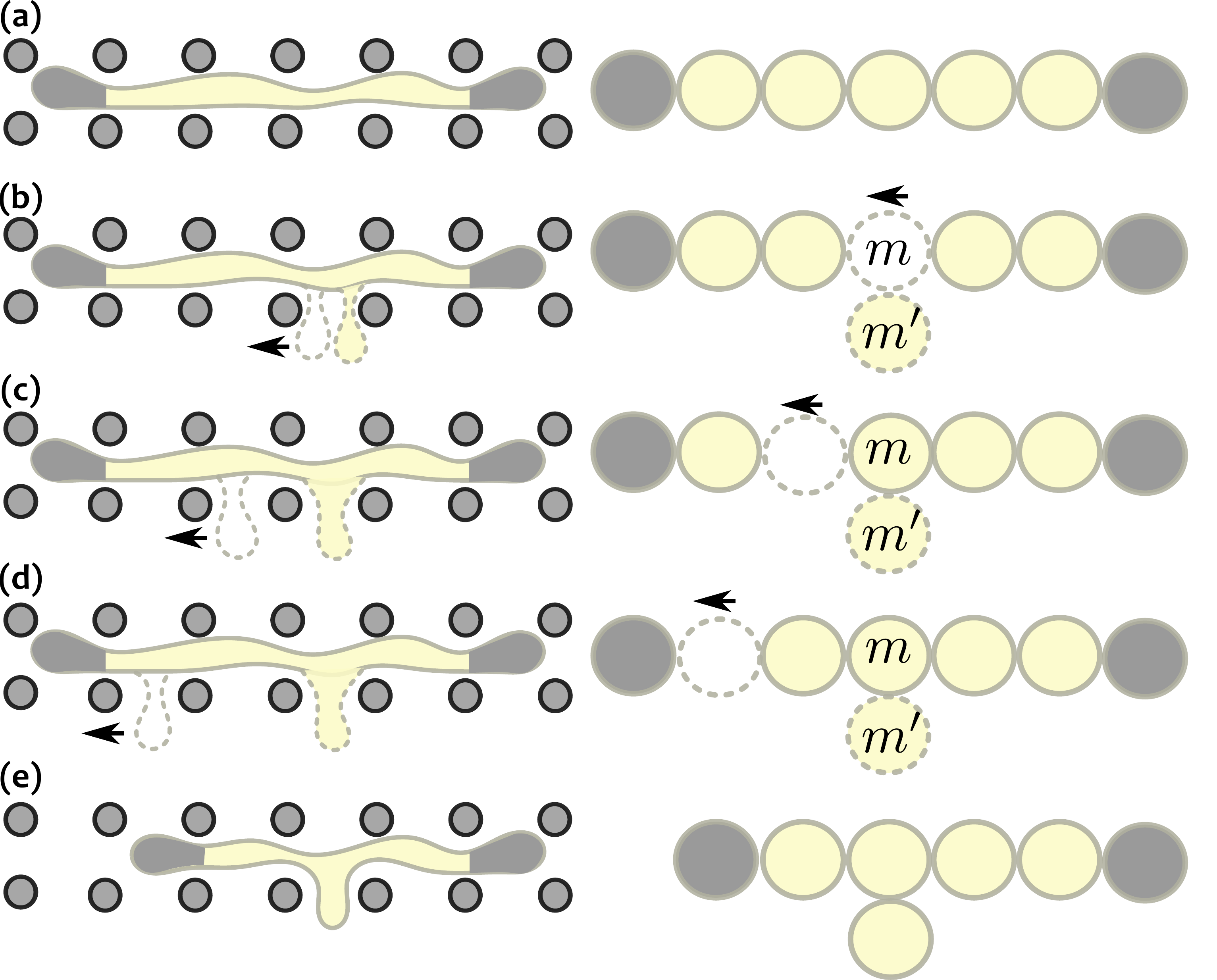}
\caption{Sketch of the ``kink''-``anti-kink'' dynamics. (a) A pair ``kink''-``anti-kink'' is created at $m$. The kink settles and becomes $m^\prime$, while the anti-kink starts a random walk from $m$. (b)-(c) The ``anti-kink'' travels to the left departing from the kink. (d) The ``anti-kink'' hits a free end (shaded in grey) and annihilates, generating the configuration in (e). (e) New configuration generated by the algorithm, where the old free end to the left is removed and substituted by a new free end.}
\label{fig:AntiKink}
\end{figure}

\subsection{``Kink''-``Anti-Kink'' Dynamics}
This algorithm is a novel way of dealing with long-ranged correlations introduced by the fact the chains are closed, and therefore have a well-defined topological state. Implementing these topological constraints is essential in simulating entangled ring polymers. The hindering of the motion when the terminal segments are threaded by another segment, \emph{i.e.} forbidden to retract (Fig. \ref{fig:Hop}(e)) is taken into account via the following set of rules (one can visualise these in Fig.~\ref{fig:AntiKink} and Fig.~\ref{fig:Hop}(e)). Starting from the configuration in Fig.~\ref{fig:AntiKink}(a), when a move $m\rightarrow m^\prime$ is energetically accepted, a pair ``kink''-``anti-kink'' (or mass-hole) is generated at site $m$ (Fig. \ref{fig:AntiKink}(b)). While the kink becomes instantaneously a new  virtual segment $m^\prime$, the anti-kink starts a random walk from $m$ which can either hit (\emph{i}) $m^\prime$, in which case the configuration goes back to the one in Fig. \ref{fig:AntiKink}(a), (\emph{ii}) a terminal end which is free (Fig. \ref{fig:AntiKink}(d)), in which case the end is retracted and the new virtual segment $m^\prime$ becomes part of the chain which becomes the one in Fig. \ref{fig:AntiKink}(e) or (\emph{iii}) a terminal end which is pinned, in which case the end is not retracted, the move is rejected and the configuration goes back to the initial state (case shown in Fig. \ref{fig:Hop}(e)). This algorithm allows for rejected moves caused by long ranged constraints introduced by the fact that the chains are closed and have to preserve their topological state. In other words, one can see this algorithm as describing elastic deformation of the chains which can protrude from any point along their contour, as opposite to linear polymers which have to free their ends before they can relax their backbone. This elastic deformation introduces a displacement of mass which can be described as a pair mass-hole or ``kink''-``anti-kink'. 
While the kink describes the protrusion/extension attempted by the chain, the anti-kink probes the ``availability'' of ends which can be retracted, or free (not pinned) ends. This procedure represents a novel way of testing the entanglement of the chain. It has the advantage that it can test long-ranged constraints such as those represented by penetrations and hence it is sensitive to self-entanglements. In fact, the larger the number of pinned ends the more likely it is that a move is rejected. 

This algorithm becomes identical to de Gennes' kink-gas diffusion in the limit $p_{th}=0$, \emph{i.e.} when no threadings are allowed. On the other hand, in the case $p_{th}>0$, we will see that this algorithm produces profoundly different behaviour which sheds light on the properties of self-entangled ring polymers in a gel.
At every Monte-Carlo time-step, every bead in the system is considered in turn, on average. By using this algorithm, we simulate chains which can diffuse by extension/retraction of their segments. The retraction is constrained by the presence of self-threading segments which hinder the chain slithering. We implicitly assume that the relaxation of a single anti-kink is much faster than the extension of a new segment. The consequence of this is twofold: (1) only one ``anti-kink'' (or ``hole'') at the time is allowed to travel along the chain and (2) the time scale at which the motion of the chains takes place is the relaxation time of the ``anti-kinks''. In other words, we reproduce the amoeba-like diffusion of the polymers at time scales larger than the ``anti-kinks'' diffusion. This choice was made to give more emphasis on the long-time behaviour of the polymer dynamics. 

It is also interesting to notice that our model naturally maps to a model for annealed branched polymers \cite{Iyer2012}.
For $p_{th}=0$ this model maps to the well-established bond-fluctuation model \cite{Carmesin1988}, where the set of allowed bonds are restricted to preserve the topological state of the rings. From another point of view, one can notice that by setting $p_{th}=0$, we forbid the presence of loops in the configuration of the animals. In this case, our model is equivalent to a blob picture\cite{Obukhov1994} for the double-folded rings, where the entanglement length is fixed to one lattice spacing. 
As $p_{th} \rightarrow 1$, the probability of finding loops increases. It is worth noting that only two beads are allowed on the same site at the same time. Hence, even when $p_{th}=1$, the lattice animals are never completely ideal. Also, we do not allow for configurations in which two segments are sharing the same site but are not threading. The motivation for this is the following: by coarse-graining the entire unit cell to one bead, we lose information on the local configuration inside each unit cell and therefore we cannot tell whether two chains sharing the same cell are threading or not. We arbitrarily choose to always label them as ``threading''. Making this choice over-counts the number of self-threadings, however, we find that our results are independent on the details of the model and the precise value of the free parameter $p_{th}$, and clearly demonstrate the importance of self-threading constraints on the motion of long ring polymers in a gel.
\begin{figure*}[t!]
\centering
\includegraphics[scale=0.7]{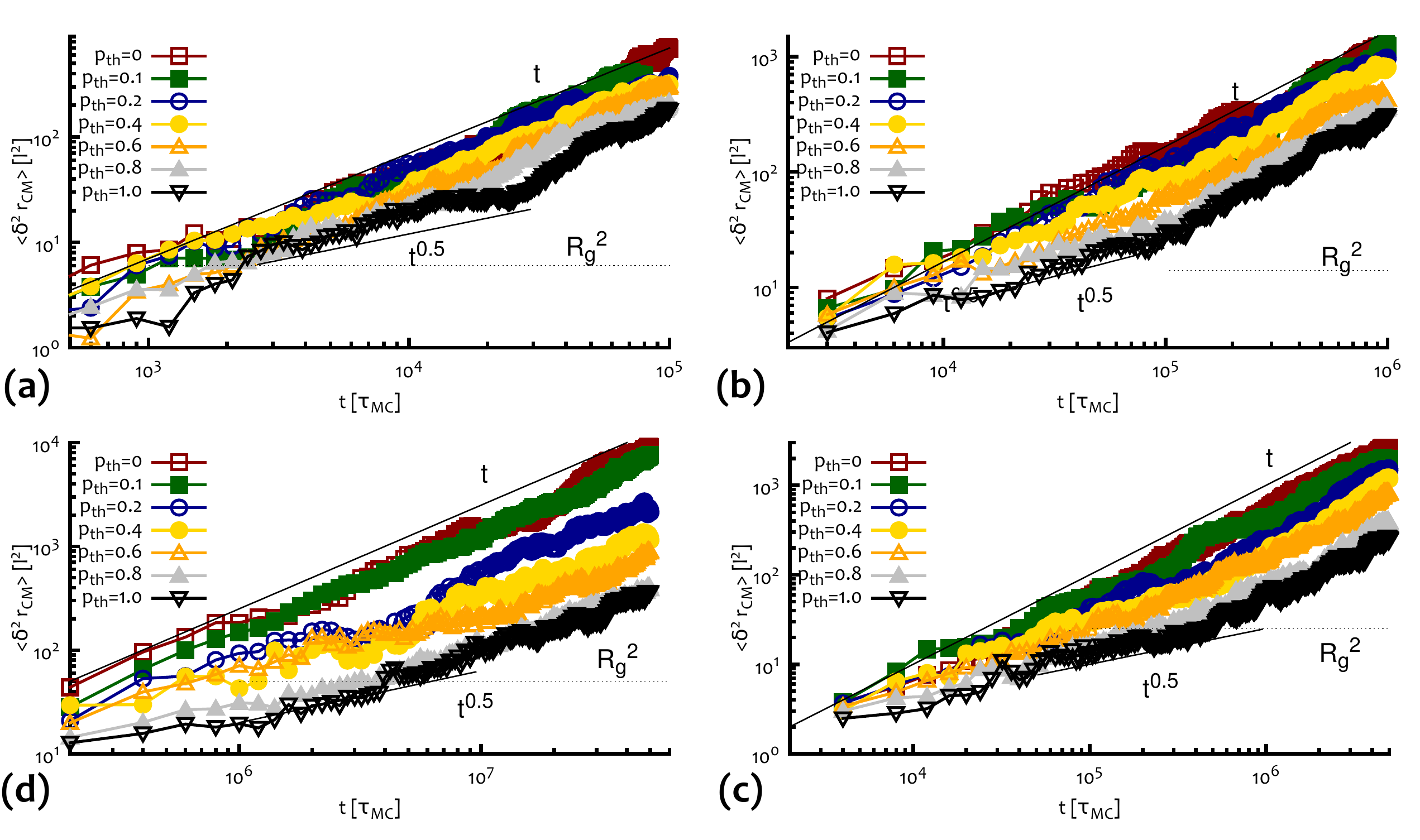}
\caption{(Colour online). Mean square displacement of the centre of mass of the rings as a function of the time for different values of the threading probability $p_{th}$ and increasing length $M$. We observe sub-diffusive behaviour $\delta^2 r_{CM} \sim t^{x}$ with $x<1$ \blue{until intermediate times}, even at moderate probability of threading $p_{th}$. The slowing down can only be caused by an increasing number of self-threadings, which become more important for longer rings (see text for details). In clockwise order: (a) $M=32$, (b) $M=64$, (c) $M=128$ and (d) $M=256$ beads. \blue{(see text for details)}.}
\label{fig:MSD}
\end{figure*}\\

\section{Results}
\subsection{Diffusion Coefficient}
By tuning the free parameter $p_{th}$ we can study the effect of self-threading on the polymer's motion. In the case $p_{th}=0$ we expect that the chains follow a pure amoeba-like diffusion \cite{Cates1986}, where the diffusion of ``anti-kinks'' takes a time of order $\tau_{kink} \sim M^2$ to travel a distance $R_g$. Since at every time step, all $M$ segment attempts to move, on average, we expect that the time taken for the centre of mass to diffuse one $R_g$ scales as
\begin{equation}
T_r(p_{th}=0) \equiv T_0 = M \tau_{kink} \sim M^{3}
\end{equation}
and consequently, the diffusion coefficient of the centre of mass of an isolated chain in gel is:
\begin{equation}
D_{CM}(p_{th}=0) \equiv D_0 = \dfrac{R^2_g}{T_0} \sim M^{2\nu - 3} = M^{-2}
\label{eq:Dcm}
\end{equation}
since $\nu = 1/2$ in 3d for self-avoiding rings in gel \cite{Parisi1981}. The results are incidentally the same as for reptating linear polymers, as obtained previously \cite{Cates1986}. This is due to the fact that the exponent $\nu$ for self-avoiding lattice animals coincides with the value for Gaussian chains. We measured the mean square displacement of the centre of mass $\langle \delta^2 r_{CM} \rangle$ as a function of time and for different values of the chains' length and probability of threading (see Fig.~\ref{fig:MSD}). Since the model does not capture the dynamics at time-scales shorter than the kink relaxation, we expect to observe pure free diffusion of the ring's centre of mass. In fact, for $p_{th}=0$, a free diffusive behaviour throughout the time window is obtained. On the other hand, for $p_{th} \rightarrow 1$ the mean square displacement of the centre of mass shows sub-diffusive behaviour at intermediate times with crossover to free diffusion only at longer times. This behaviour is unambiguously related to the presence of self-threadings, since the system is in the dilute regime. In other words, allowing the rings to self-thread results in an increasingly important self-constraint on the motion. These contributions on the motion are ultimately caused by the preservation of the topological state of the rings, which have to stay unknotted and unlinked from both the gel and themselves. \blue{The length-scales associated with the crossover from sub-diffusive behaviour to the free diffusive one lie between $(R^2_g)^{1/2}$ and $(10 R^2_g)^{1/2}$. This feature is in agreement with previous findings in systems of rings with similar topological constraints \cite{Michieletto2014,Halverson2011a}. We argue that these length-scales are related with the loss of threadings, and that the rings have to travel many times their own average size before relaxing all the threadings, \emph{i.e.} freely diffusing.} We expect that linear or branched polymers would not undergo the same change in diffusion by allowing sites with double occupancy, as there is no defined topological state to be conserved. 

\begin{figure}[t]
\includegraphics[scale=0.7]{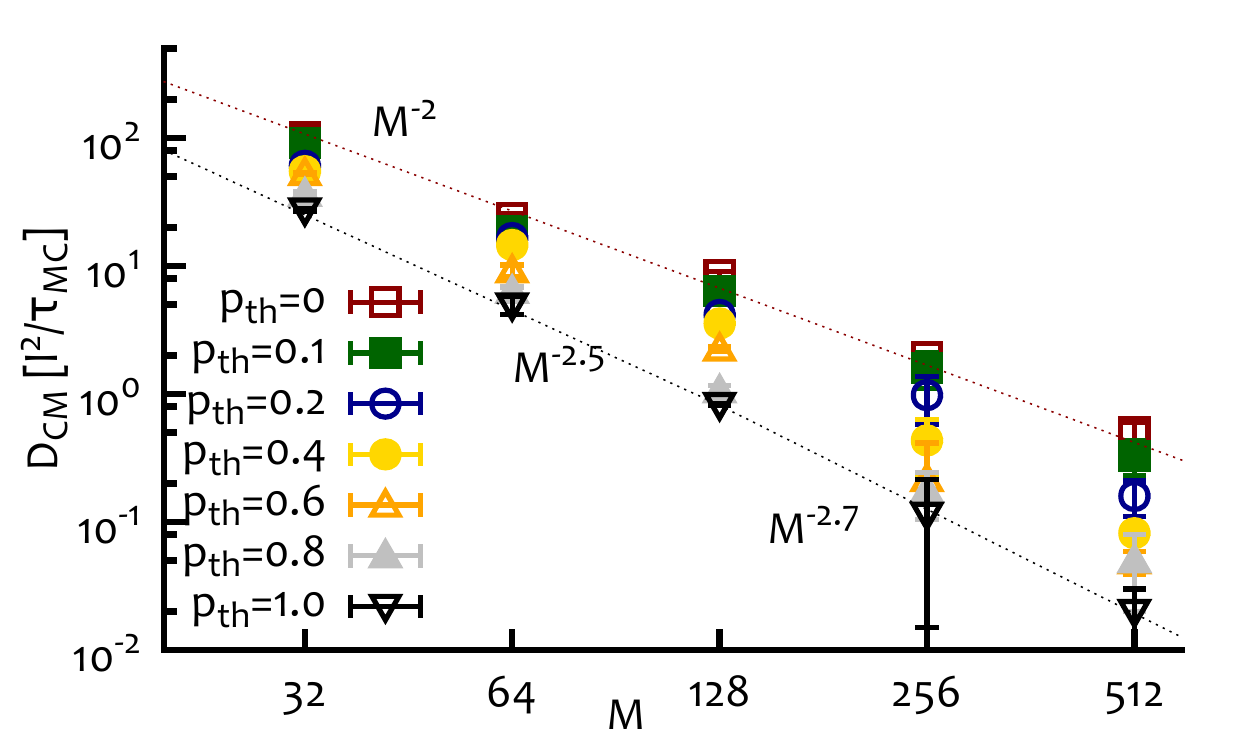}
\caption{(Colour online). Log-log plot showing the diffusion coefficient of the centre of mass $D_{CM}$ computed as $\lim_{t \rightarrow \infty} \delta^2 r_{CM}/6t$ as a function of the chains length $M$. One can observe pure amoeba-like (free) diffusion at $p_{th} = 0$, which scales as $D_{CM} \sim M^{-2}$ according to eq. \eqref{eq:Dcm}, or slower diffusion $D_{CM} \sim M^{-\alpha}$ with $\alpha>2$ for increasing $p_{th}$.}
\label{fig:DvsM}
\end{figure}

In Fig.~\ref{fig:DvsM} we show the scaling behaviour of the diffusion coefficient of the centre of mass $D_{CM}(p_{th})$. Notice that $D_{CM}(0) \equiv D_0 \sim M^{-2}$, reproducing the scaling regime obtained for amoeba-like motion, as in previous works~\cite{Rubinstein1986,Cates1986,Obukhov1994}. For higher values of $p_{th}$ the rings diffuse slower, due to the chains' self-threading. The $M$ dependence of the diffusion coefficient of the centre of mass of the rings is observed to become more severe as $p_{th}\rightarrow 1$. 

\subsection{Radius of Gyration}

\begin{figure}[t]
\includegraphics[scale=0.7]{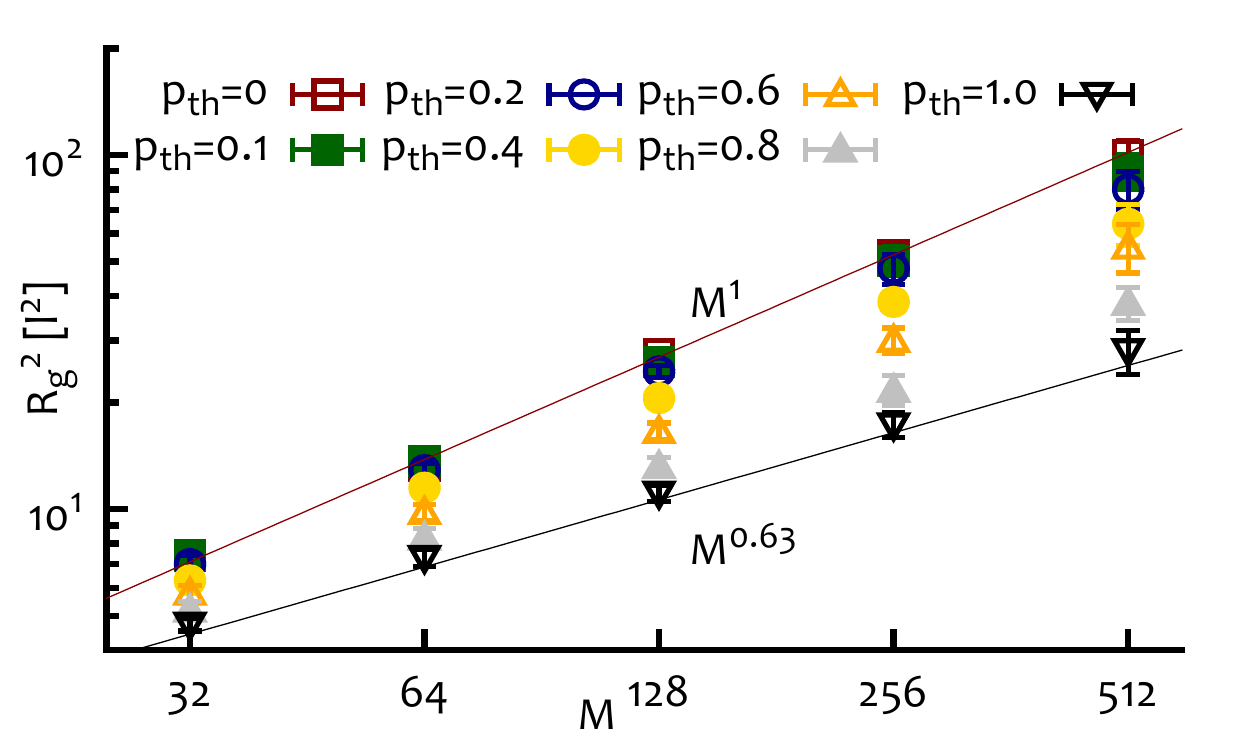}
\caption{(Colour online). Log-log plot showing the radius of gyration of the chains as a function of the rings' length $M$. The prediction $R_g \sim M^{1/2}$ for ring polymers in gel is obtained at $p_{th} \rightarrow 0$. For $p_{th} \rightarrow 1$ we observe $R_g \rightarrow M^{1/d}$, as predicted is previous works~\cite{Daoud1981}. }
\label{fig:RGvsM}
\end{figure}
\begin{figure}[h!]
\includegraphics[scale=0.7]{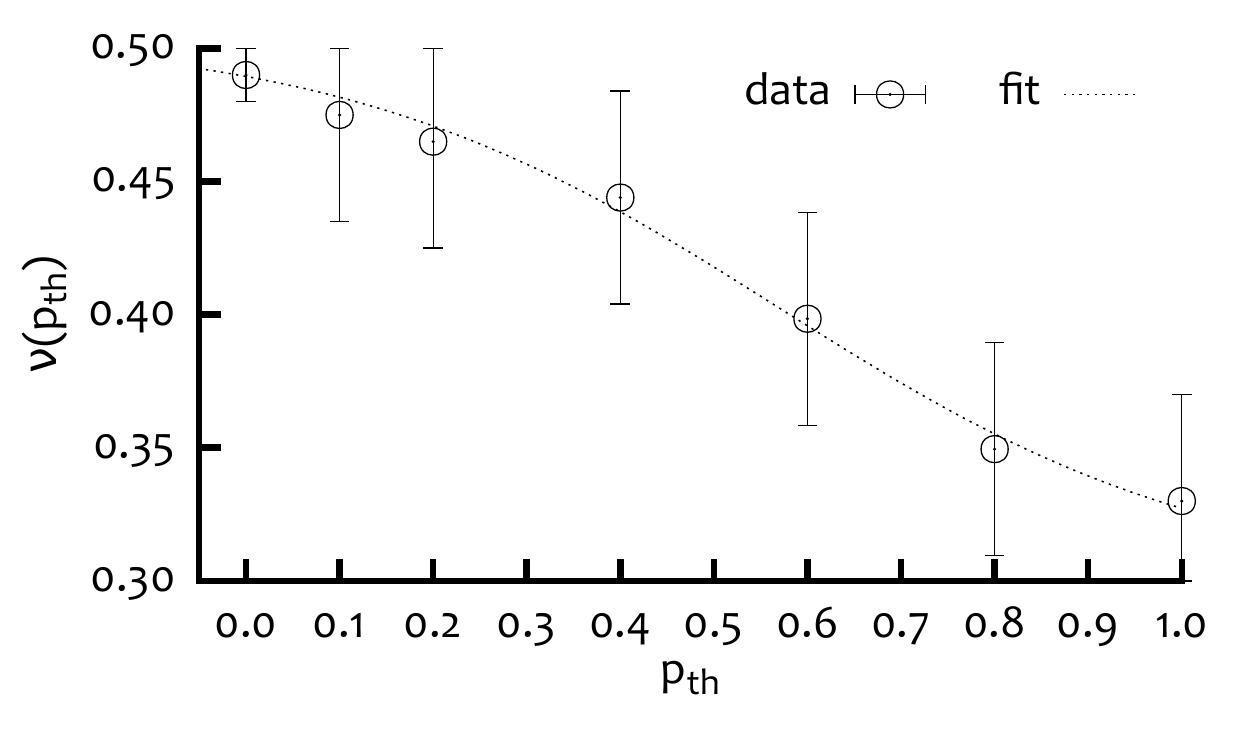}
\vspace*{-0.3 cm}
\caption{Values of the exponent $\nu$ as a function of $p_{th}$. The fit has been performed using a sigmoid function $\nu(x) \sim A/(1+e^{w(x-p_{th}^*)})+B$. The result suggests a continuous transition from lattice animal behaviour ($\nu=1/2$) to fractal globule behaviour ($\nu =1/3$) as $p_{th}$ crosses $p_{th}^* \sim 0.5$.}
\label{fig:Nupth}
\end{figure}

The radius of gyration at $p_{th}=0$ is in agreement with the mean-field prediction ($\langle R^2_g \rangle \sim M$) for lattice animals with excluded volume~\cite{Lubensky1979,Parisi1981,Daoud1981}. Higher values of $p_{th}$ allow the polymers to partially self-overlap (only a maximum of two beads per site are allowed) and therefore reduce their size. The mean-field prediction for ideal lattice animals~\cite{Lubensky1979} ($R_g \sim M^{1/4}$) breaks down at $d<d_c$ with $d_c=8$ for lattice animals in good solvent, hence we do not expect this to be valid in the present work. The scaling regime at $p_{th}=1$ resembles instead a regime in which the lattice animals have screened out two-body excluded volume interactions, \emph{i.e.} the repulsive second virial coefficient is zero, while retaining three body excluded volume. 
It has been recently shown that the gyration radius of rings in the melt assume the minimum value allowed compatible with such excluded volume constraints, \emph{i.e.} a fractal globule conformation with $\langle R_g \rangle \sim M^{1/3}$ in $d=3$~\cite{Grosberg2013,Rosa2013}, which is compatible with our findings (see Fig.~\ref{fig:RGvsM}). The values of the entropic exponent $\nu$ observed in the Kinetic Monte-Carlo simulations are also in agreement with the value obtained via the Molecular Dynamics simulation (see S.I.).

We also investigate the functional dependence of the exponent $\nu$ on the value of the free parameter $p_{th}$ (see Fig.~\ref{fig:Nupth}). The function that best fits the data is a sigmoid function which continuously crossovers from a lattice animal value of $\nu =1/2$ to a fractal globule value of $\nu=1/3$ as $p_{th}$ crosses $p_{th}^*\simeq0.5$. Our model captures such crossover by just varying the free parameter of the model. Under this perspective, $p_{th}$ plays the role of an effective second virial coefficient, which regulates two body repulsion. This also suggests that steric effects are intimately related to the hindering of the  motion, since the topological state of the rings has to be preserved, \emph{i.e.} the threading segments cannot cross. We also want to stress that for any polymer with a non-closed shape embedded in a background gel, a variation of the second virial coefficient is not expected to affect its motion as severely as in the case of rings, since it is the preservation of the topological state at the heart of the constraint on the motion. In light of this, in the next section we study the statistics of self-threadings and their functional dependence on length $M$ and probability of threading $p_{th}$.

\subsection{Self-Threadings and Pinned Ends}

\begin{figure}[t]
\includegraphics[scale=0.7]{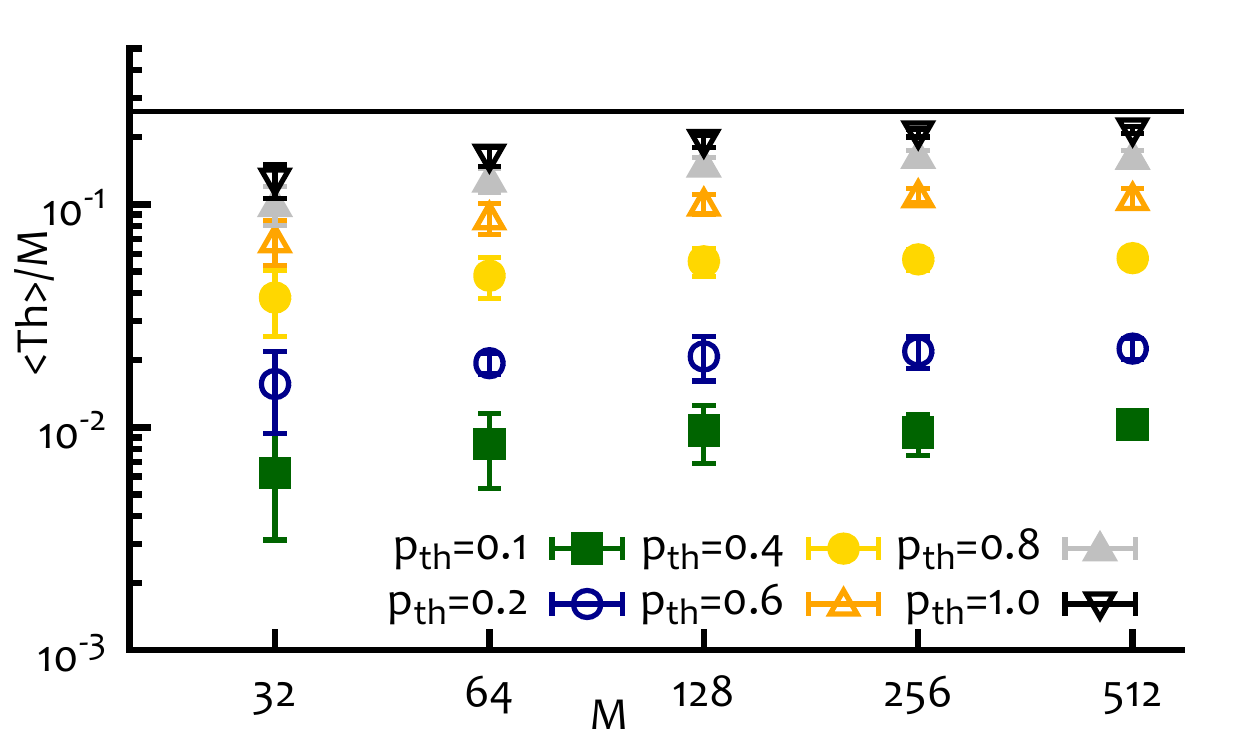}
\caption{ (Colour online). Log-log plot showing $\langle Th \rangle/M$ as a function of chains' length $M$. For large $M$, the number of threadings per chain scales extensively in $M$.}
\label{fig:THvsPth}
\end{figure}

In Fig.~\ref{fig:THvsPth} we show that the number of threadings per chain is found to scale extensively with the number of beads $M$ times a constant that in general can depend on $p_{th}$, \textit{i.e.}  $\langle Th \rangle \sim A(p_{th}) M$. As one can see from Fig.\ref{fig:Hop}, a self-threading looks like a loop in the polymer conformation. Previous work confirmed that the number of loops in lattice animal conformations does not represent a critical quantity~\cite{Lubensky1979,Daoud1981}, \emph{i.e.} the number remains constant as the length of the rings increases. Here we want to stress that, in our model, self-threadings are not completely equivalent to loops. One can see this by looking at Fig. \ref{fig:Hop}(e). As one segment threads through another, we do not change the functional unit of the interacting beads, \emph{e.g.} the two ends threading through each other in the figure  remain distinct ends which, \magenta{computationally, is implemented by stacking the beads on top of each other}.
In fact, as shown in Fig. \ref{fig:THvsPth}, for every value of $p_{th}>0$ we obtain $\langle Th \rangle \sim M$ in the large $M$ limit. This result clearly states that an abundant number of self-threading will emerge as the length of the rings increases, ultimately hindering the motion of the rings by pinning more and more ends. \\

\begin{figure}[t]
\includegraphics[scale=0.7]{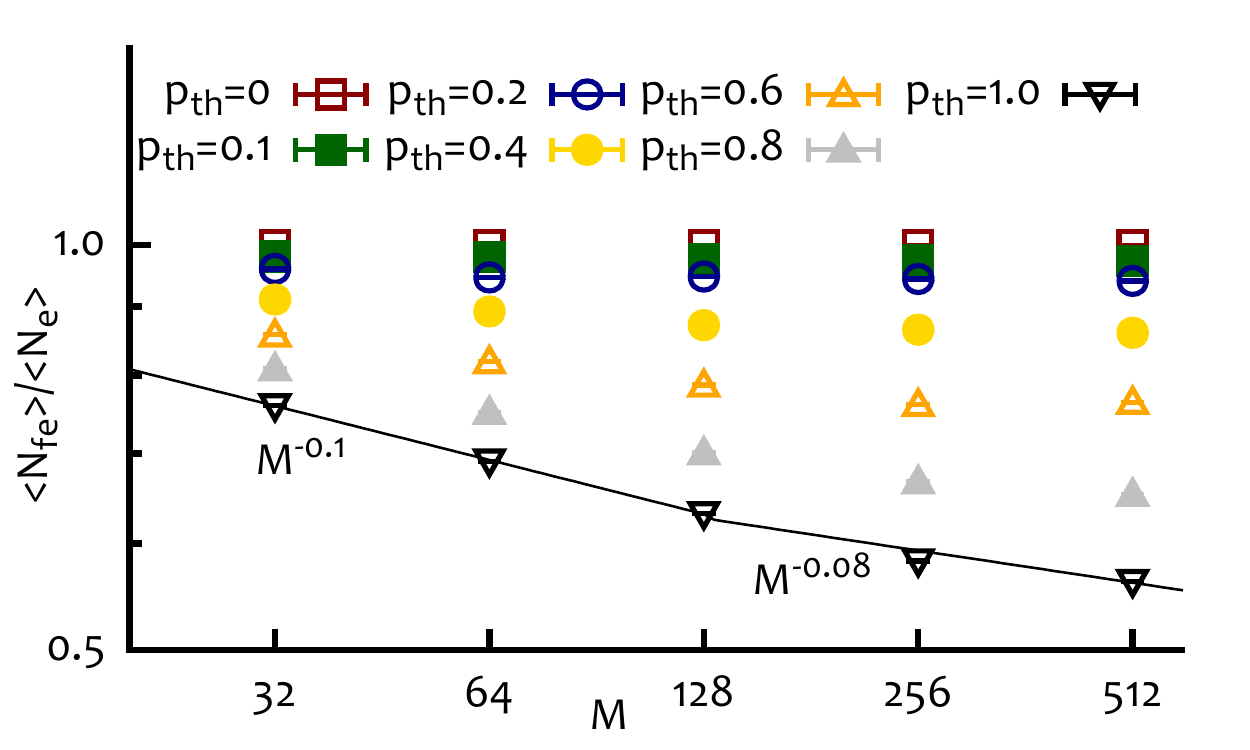}
\caption{(Color online). Log-log plot showing the ratio $\langle N_{fe}\rangle/\langle N_e \rangle$ as a function of the chains' length. The curves seem to approach the limiting value of $1/2$, at which half of the ends would be pinned and cannot contribute to the motion. }
\label{fig:FENDSoverENDSvsM}
\end{figure}

In Fig. \ref{fig:FENDSoverENDSvsM} we show the equilibrium statistics of the fraction of free ends (\emph{i.e.} ends that are not pinned and can contribute to the rings' motion) over the total number of ends in a lattice animal. The number of free and pinned ends are related by the fact that the sum of them must be equal to the total number of ends, \emph{i.e.} $\langle N_e \rangle = \langle N_{fe} \rangle + \langle N_{pe}\rangle$. We observed (data not shown) that the number of ends  scales linearly with the length $M$. Of these, a fraction $\langle N_{fe}\rangle/\langle N_e \rangle$ are free to retract, while a fraction $1-\langle N_{fe}\rangle/\langle N_e \rangle$ are threaded, or pinned, and hence not free to retract. In Fig. \ref{fig:FENDSoverENDSvsM} we show that as the length of the rings increases, the fraction of free ends decreases and this effect is more evident as $p_{th}$ is closer to $1$. As the number of ends that can contribute to the motion becomes smaller, we expect the dynamics to become slower, too. In the next section we study how the presence of self-threadings affects the relaxation of the chains.

\subsection{Relaxation Dynamics}

\begin{figure}[t]
\includegraphics[scale=0.7]{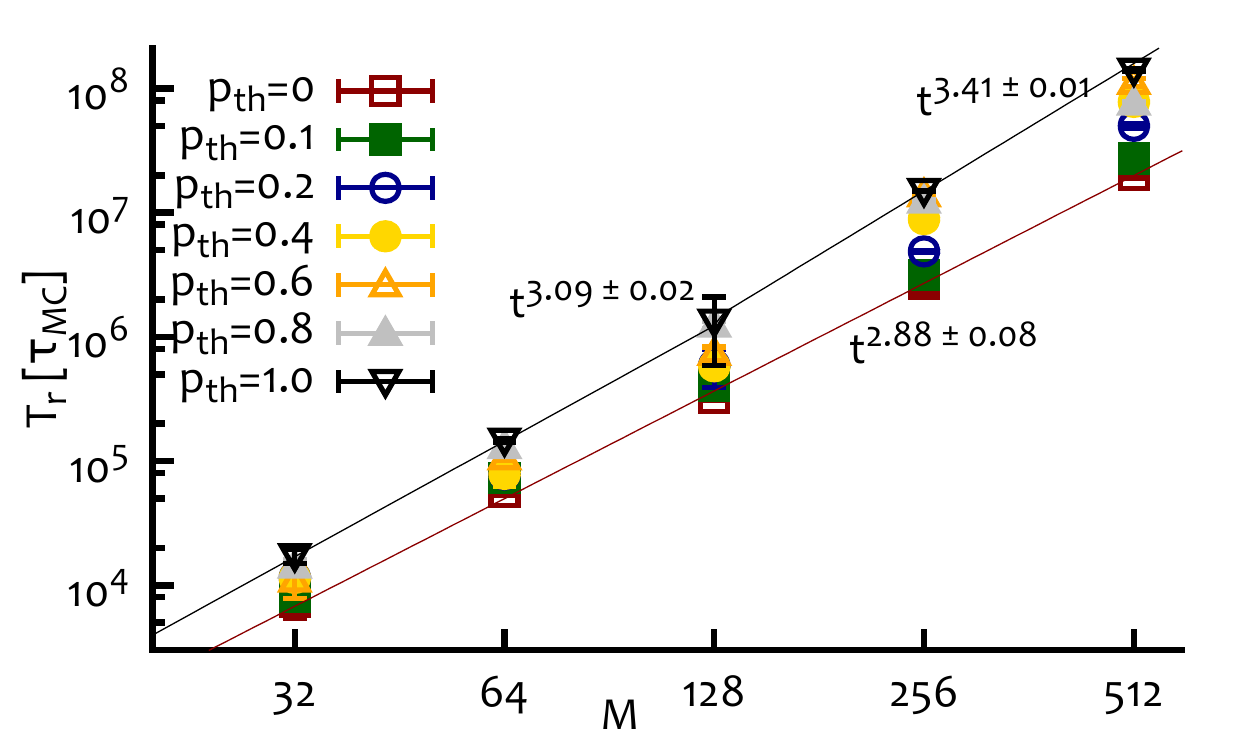}
\caption{(Colour online). Log-log plot showing the relaxation time $T_r = \langle R^2_g \rangle/ D_{CM}$ as a function of the rings length $M$. The prediction for amoeba-like motion is obtained at $p_{th}=0$. For higher values of $p_{th}$ we observe a slowing down of the chain relaxation.}
\label{fig:TRvsM}
\end{figure}
\begin{figure}[t]
\includegraphics[scale=0.7]{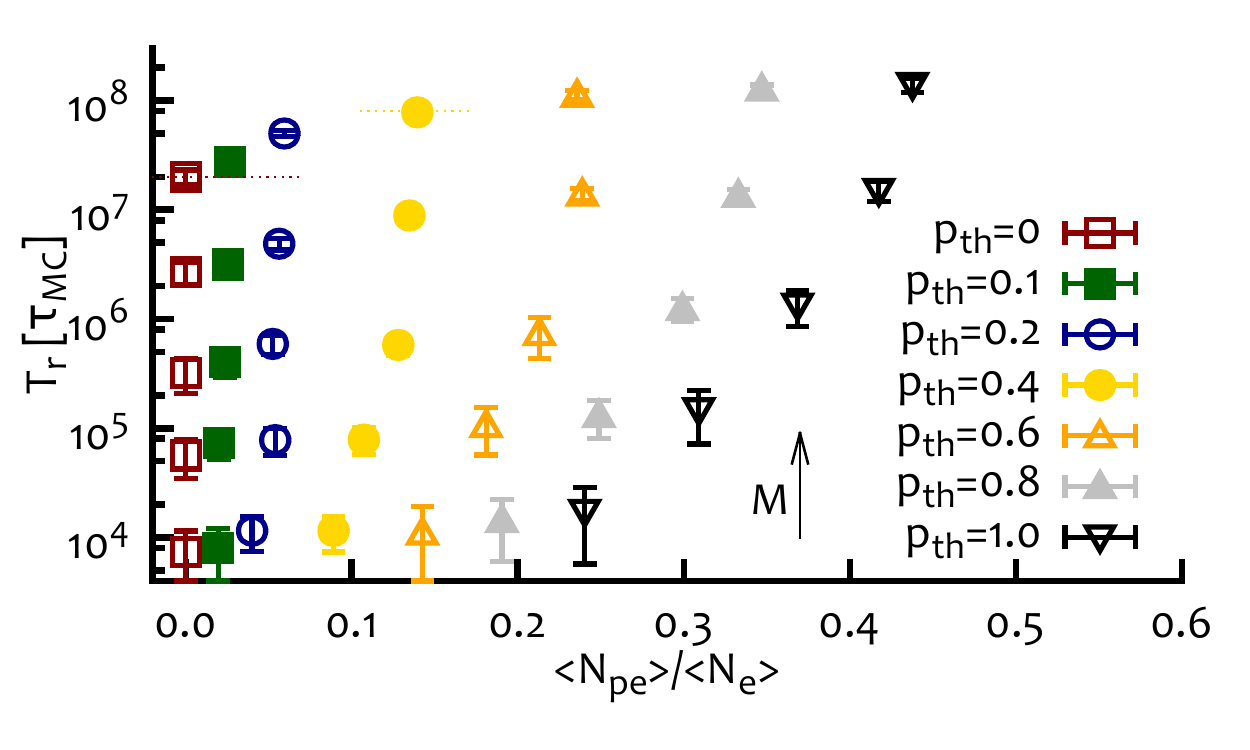}
\vspace*{-0.3 cm}
\caption{(Color online). Log-linear plot showing the relaxation time $T_r$ as a function of the fraction of pinned ends $\langle N_{pe}\rangle/\langle N_e \rangle$. The length of the rings $M$ increases upwards. Notice that even for moderate values of $p_{th}$ we obtained substantial slowing down. For $M=512$ the relaxation time $T_r$ at $p_{th}=0.4$ (yellow dotted line) is five times larger than $T_r$ at $p_{th}=0$ (red dotted line).}
\label{fig:TRvsPENDS}
\end{figure}

The slowing down due to self-threading of the chain is apparent from the plot of the relaxation time of the chain (see Fig. \ref{fig:TRvsM}). This increases dramatically with chain size, with a power law which appears to depend on $p_{th}$. In Fig. \ref{fig:TRvsPENDS} we show the relaxation time $T_r$ as a function of the fraction of pinned ends $ \langle N_{pe}\rangle/\langle N_e \rangle = 1-\langle N_{fe}\rangle/\langle N_e \rangle$. One can observe that, compared to the value at $p_{th}=0$, the relaxation time at $p_{th}>0$ can be substantially larger. For $M=512$, even at moderate values of $p_{th}$ one can observe a significant slowing down, for instance in Fig.~\ref{fig:TRvsPENDS} we show that $T_r(p_{th}=0.4)$ (yellow dotted line) is roughly five times larger than the relaxation time at $p_{th}=0$ (red dotted line), which could easily be observed experimentally. 

\begin{figure}[t]
\includegraphics[scale=0.7]{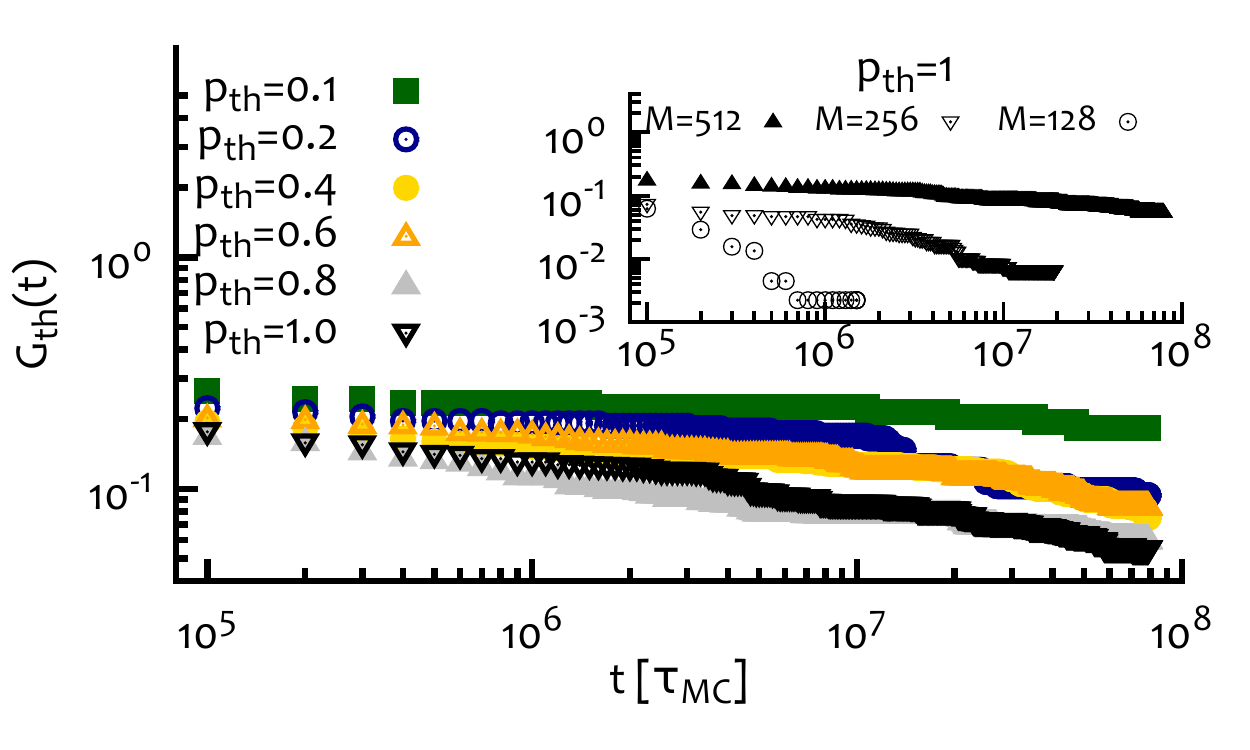}
\caption{(color online). Fraction of threadings present $t$ time-steps after that $p_{th}$ is turned off for $M=512$ and different values of $p_{th}$ used to bring the system to equilibrium. $G_{th}(t)$ shows a slow decay that lasts for several decades. This suggests the presence of a hierarchical structure of self-threadings and long-lived correlations affecting the long time dynamics. \red{For the longest rings all the threadings did not disappear in the simulation runtimes accessible to us.} (Inset) $G_{th}(t)$ for $p_{th}=1$ and different values of $M$.}
\label{fig:ThRelax}
\end{figure}

\begin{figure}[t]
\includegraphics[scale=0.7]{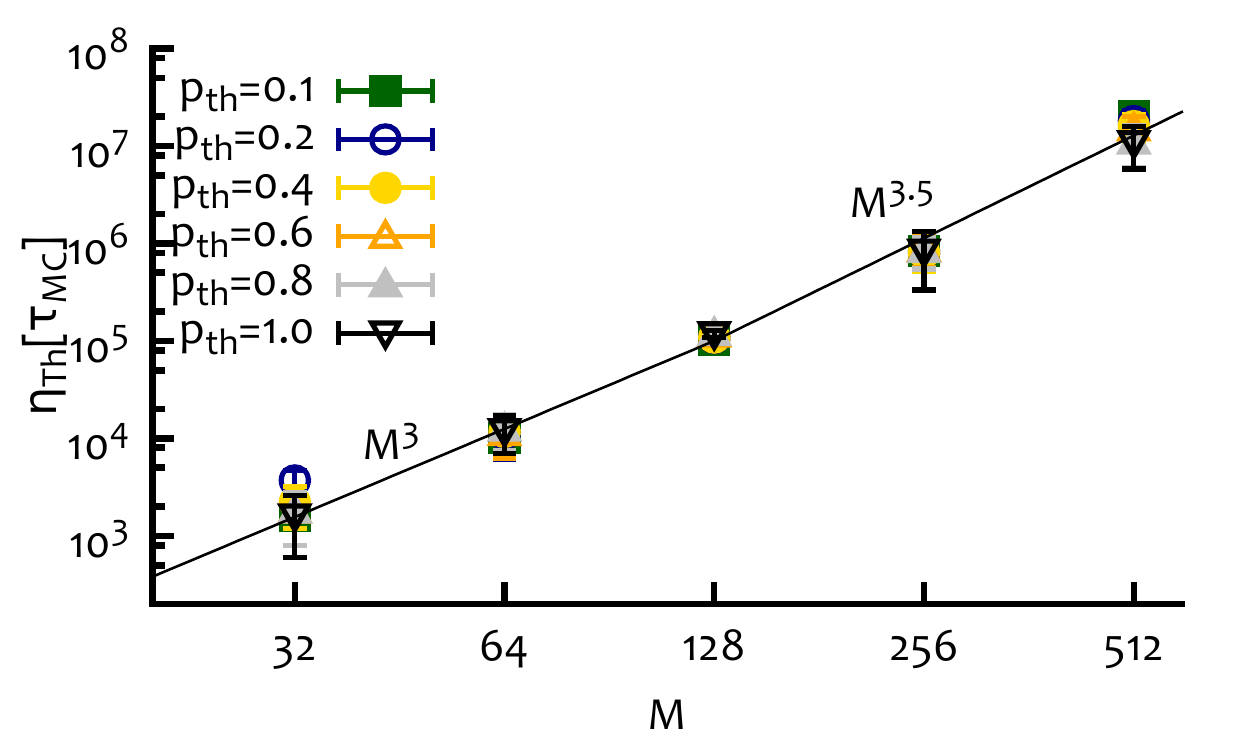}
\caption{(color online). Numeric integral of $G_{th}(t)$, $\eta_{th} = \int_0^{\infty} G(t) dt$. Independently on the value of $p_{th}$, the data points are fitted by the power law $\tau_{th} \sim M^\alpha$, with $\alpha = 3$ for $M<128$ and $\alpha=3.5$ for $M>128$. \red{The arrow at M=512 indicates that the values of $\eta_{Th}$ for the longest rings represent only a lower bound as we could not observe the removal of all the threadings within the simulation runtime.} }
\label{fig:tauThRelax}
\end{figure}

Finally, we study the \red{resistance} of the threadings by taking fully equilibrated configurations with $p_{th}>0$, and turning the free parameter $p_{th}$ to zero. In other words, we forbid the creation of new pinned sites in order to study the time-scale required for the self-threadings to relax. We report our findings in Fig.~\ref{fig:ThRelax} and \ref{fig:tauThRelax}. In Fig. \ref{fig:ThRelax} we plot the threading relaxation function $G_{th}(t)$ as a function of time $t$. This represents the average fraction of self-threadings present $t$ time-steps after we set $p_{th}=0$. We argue that if the self-threadings were uncorrelated with each-other, we would expect a exponential decay of $G_{th}(t)$ from $G_{th}(t=0)$, \red{which equals one by definition of $G_{th}$}, to zero. On the contrary, after an initial drop, we observe \red{a very slow power law decay} on the time-scales comparable with the chains longest relaxation times. Such behaviour is quite reminiscent of the reptating mechanism for linear chains, where the entanglement with the tube has to be removed one segment at a time starting from the ends of the chain. We argue that in our model, while the chains can diffuse in an amoeba-like fashion in space, \emph{i.e.} by generating new protrusions at any point along their backbones, they are forced to undergo a process that it is more similar to reptation, \emph{i.e.} diffusion along backbone, to release self-threadings. This suggests that self-threadings create a nested network of constraints that is significant on the time-scales of the rings motion. \red{For the longest rings, we could not observe the removal of all the threadings, even at very long times (see Fig.\ref{fig:ThRelax}).}

The difference between rings in gel and linear polymers is that new ends can be formed everywhere along the rings contour, so that no real tube can confine the rings diffusion in space. Nonetheless, self-threadings have to be removed in an hierarchical way for the chains to diffuse freely. As shown in Fig. \ref{fig:ThRelax}, this process can take a time comparable to the longest relaxation time of the chains.
The equivalent of the zero-shear viscosity for the threadings is computed as the numerical integral of $G_{th}(t)$, \emph{i.e.} $\eta_{th} = \int_0^{\infty} G(t) dt$ (see Fig. \ref{fig:tauThRelax}).  Its value seems to be weakly dependent on the value of $p_{th}$ and to scale as $\eta_{th}\sim M^\alpha$, with $\alpha \geq 3$. \red{For the longest rings, we could compute only the lower bound of $\eta_{th}$, as the curves in Fig. \ref{fig:ThRelax} do not decay to negligible values within the simulation window. This is  indicated on Fig. \ref{fig:tauThRelax} by the arrow coming out the data points at $M=512$.} This is consistent with the fact that self-threadings represent long-lived correlations on the chains motion and represent severe constraints on  the dynamics of very long chains in gel.

\section{Conclusions}
We presented a Kinetic Monte-Carlo algorithm to simulate a dilute solution of ring polymers in a gel. This algorithm reproduces the known results for rings in gel \cite{Obukhov1994, Cates1986} and adds the possibility of taking into account self-threadings. The static and dynamic properties of the rings have been studied by tuning the free parameter $p_{th}$. We observed a drastic change in the polymers' behaviour (see Fig.~\ref{fig:MSD}) as $p_{th} \rightarrow 1$. In particular, we observed a sub-diffusive behaviour of the mean square displacement of the centre of mass ($\langle \delta^2 r_{CM}(t) \rangle \sim t^x$ with $x<1$) which crossovers to free diffusion ($x=1$) only at longer times. \blue{The length-scales associated with the crossover agree with previous studies of systems of rings with similar topological constraints \cite{Michieletto2014}}. The severe slowing down observed in the polymers' dynamics is caused  by the presence of long-ranged and long-lived correlations which take the form of self-threadings (see Figs. \ref{fig:DvsM},\ref{fig:TRvsM} and \ref{fig:ThRelax}). Such constraints on the polymers' diffusion are intimately related to the fact that we are studying rings, as for any other shape (which do not include the presence of closed contours) such slowing down in the dynamics is not expected. We observe that even for moderate values of $p_{th}$ the relaxation time of the rings can be almost one order of magnitude larger than the case without threadings (see Fig.~\ref{fig:TRvsPENDS}). The number of these self-threadings is found to scale extensively with the size of the chains $M$, which ensures that these can be prolific even for small $p_{th}$ for long enough chains (see Fig.~\ref{fig:THvsPth}). This suggests that self-threadings can be relevant for a complete understanding of ring diffusion in gel. It is worth noting that we expect the effects of self-threadings on the rings dynamics to become measurable experimentally only for gels that do not contain dangling ends. Isolating the effect of slowing down caused by ``impalement'' of the rings~\cite{Viovy2000} from that caused by self-threadings can, in fact, be difficult experimentally, as very little is known about the process of impalement of rings from the gel's dangling ends  (and it is focus of future work). We argue that these self-threadings might be more easily observed by employing a micro-lithographic array of obstacles~\cite{Volkmuth1992}. It is worth reminding that one can think of $p_{th}$ as an effective second virial coefficient that regulates pair interaction. In light of this, we argue that $p_{th}$ might be tuned by acting on the temperature. In general, we expect the value of $p_{th}$ to be small, hence we conjecture that self-threadings represent a real hindering on the dynamics only of very large polymers, \textit{i.e.} of order of thousands of Kuhn segments. In terms of DNA plasmids, where a Kuhn length is $l_k \sim 100$ $nm$ we expect the self-threadings to become relevant in the dynamics for mega-base sized DNA plasmids, which can be analysed with a very sparse gel to avoid breaking the samples. Finally, we suggest that such self-threadings can also contribute in the process of ``irreversible self-trapping'' of polymers in gel \cite{Viovy1992, Duke1996, Viovy2000}. In light of our results, we argue that ring polymers can easily become irreversibly self-trapped by undergoing self-threading and then by being pulled taut around the gel's structure by an electric field (see Fig.~\ref{fig:Hop}(f)).\\

\subsubsection*{Acknowledgement \hspace*{0.1 cm}  } 
DMi acknowledges the support from the Complexity Science Doctoral Training Centre at the University of Warwick with funding provided by the EPSRC (EP/E501311). EO acknowledge financial support from the Italian ministry of education grant PRIN 2010HXAW77. We also acknowledge the support of EPSRC to DMa, EP/I034661/1, and MST, EP/1005439/1, the latter funding a Leadership Fellowship. The computing facilities were provided by the Centre for Scientific Computing of the University of Warwick with support from the Science Research Investment Fund.

\footnotesize{
\bibliography{SelfThreadingRingsInGelFIXED} 
\bibliographystyle{rsc} 
}

\normalsize
\appendix
\section{Molecular Dynamics Simulation Details}

We model the ring polymers using a standard bead-spring semi-flexible model based on the Kremer Grest \cite{Kremer1990} model. Every bead in our simulation interacts via a shifted 
Lennard-Jones potential with a cut-off $r_c = 2^{1/6} \sigma$. The gel is itself made of beads which partially overlap in order to preserve the topological status of the ring polymer, \textit{i.e.} unlinked from the the gel. The beads in the gel interact only with the beads forming the polymers via the same shifted Lennard-Jones potential. The beads forming the gel are not treated in the dynamics, meaning that the background structure is fixed and static at all times. Nearest neighbour beads along the ring polymers interact via a finitely extensible non-linear elastic (FENE) potential. The non-linear chain's flexibility is then introduced by an angular potential. 
The total intra-chain potential is therefore given by the following Hamiltonian:
\begin{align}
&H_{intra} = \sum_{i=1}^{M} \left[ U_{FENE}(i, i+1) +  \phantom{\sum_{i}^M}\right.  \notag\\
&\left. \phantom{\sum_N^M} + U_{b}(i,i+1,i+2) \right] + \sum_{i=1}^{M-1}\sum_{j=i+1}^{M} U_{LJ}(i,j) \label{eq:Hintra} 
\end{align}
where $M$ is the number of beads in the ring and the terms with $i>M$ represent those interactions needed to join the ends of the polymer in a ring fashion, \textit{i.e.} a modulo-M indexing is implicitly assumed to take into account the ring periodicity. 
Each monomer has nominal size $\sigma$ and position $\bm{r}_i$, while the distance between two monomers $i$ and $j$ is given by $d_{i,j} = | \bm{r}_i - \bm{r}_{j}|$. The finitely extensible non-linear elastic potential is of the form:
\begin{equation}
U_{FENE}(i,i+1) = -\dfrac{k}{2} R_0^2 \ln \left[ 1 - \left( \dfrac{d_{i,i+1}}{R_0}\right)^2\right]  \notag
\end{equation}
for  $d_{i,i+1} < R_0$ and $U_{FENE}(i,i+1) = \infty$, otherwise; $R_0 = 1.5$ $\sigma$, $k=30$ $\epsilon/\sigma^2$ and the thermal energy $k_BT$ is set to $\epsilon$.
The bending energy, or stiffness term, takes the standard Kratky-Porod form (discretized worm-like chain):
\begin{equation}
U_b(i,i+1,i+2) = \dfrac{k_BT \xi_p}{\sigma}\left[ 1 - \dfrac{\bm{d}_{i,i+1} \cdot \bm{d}_{i+1,i+2}}{d_{i,i+1}d_{i+1,i+2}} \right] \notag
\end{equation}
where $\xi_p$ is the persistence length of the chain which is fixed at $5$ $\sigma$. Polymers are significantly bent by thermal fluctuations at contour lengths larger than the Kuhn length $l_k = 2 \xi_p$. Here, the persistence length $\xi_p$ is always assumed to be much smaller than the total length of the chain, so that the chains resemble a flexible polymer, rather than a rigid rod. The `cut and shifted' Lennard-Jones potential takes the following form:
\begin{equation}
U_{LJ}(i,j) = 4 \epsilon \left[ \left(\dfrac{\sigma}{d_{i,j}}\right)^{12} - \left(\dfrac{\sigma}{d_{i,j}}\right)^6 + 1/4\right] \notag
\end{equation} 
for $d_{i,j} < 2^{1/6}$ $\sigma$ and $U_{LJ}(i,j) = 0$, otherwise. The same potential is also used to regulate all the pair interactions between monomers belonging to the chain and the fixed mesh. The chain-mesh Hamiltonian is:
\begin{equation}
H_{chain-mesh} = \sum_{k=1}^{M_{gel}} \sum_{i=1}^{M} U_{LJ}(k,i)
\end{equation}
the index $i$ runs over the beads in the chain and $k$ runs over the beads forming the mesh. The bead mass is $m$ and the friction acting on each bead is set to $\xi / m = \tau_{LJ}^{-1}$. The integration is performed in the over-damped limit of the Langevin equation using Verlet algorithm with time step $\Delta t = 0.01$ $\tau_{LJ}$, as in previous works \cite{Halverson2011c,Halverson2011a} . 
The ring is initialised outside the mesh in a easily parametrisable fashion. Initially, a short run is performed where instead of the Lennard-Jones potential, we employ a soft repulsive potential between bonded beads with energy $E_s = 40 \epsilon$ and cutoff $r_s = 2^{1/6}$ $\sigma$. In this way we gently push the bonded monomers apart and avoid numerical blow ups. After this short run, the soft potential is replaced by the Lennard-Jones potential described above. At this point we drag the polymer inside the mesh by applying an external force on some of the monomers. Once the ring is completely contained in the gel we adapt the simulation box in order to perfectly fit the size of the gel and test the ring topology. If the ring is in a unknotted state we proceed with the equilibration run.

\section{Equilibrium Configuration}
\begin{figure}[!h]
\centering
\includegraphics[scale=0.17]{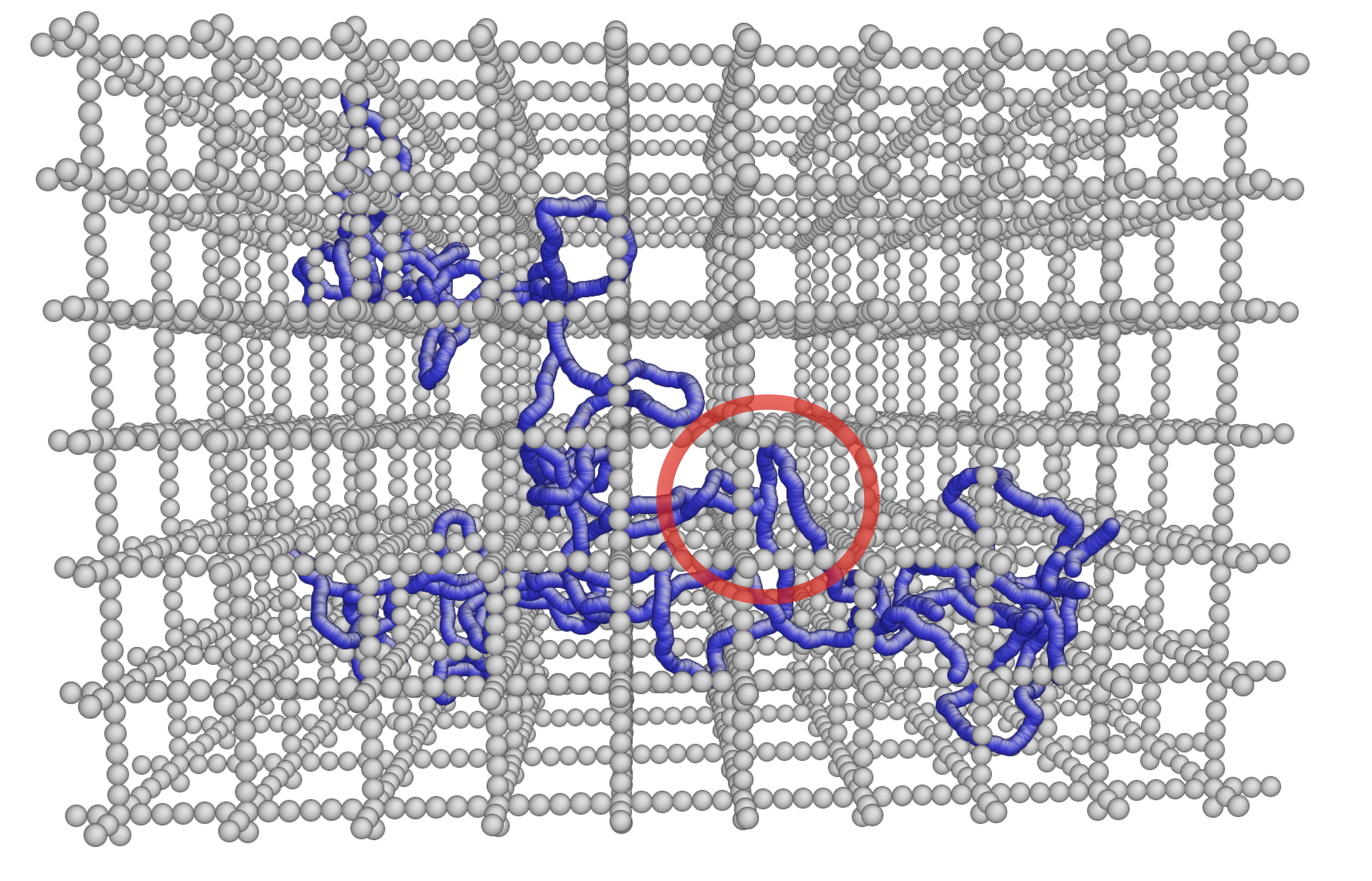}
\caption{Equilibrium configuration of the ring polymer in gel. The shape resembles that of a lattice animal. Highlighted in red, two segments of the chain which are about to self-thread. The gel structure is here thinned for clarity.}
\label{fig:LA}
\end{figure}

The presence of the gel with lattice spacing comparable with the polymer Kuhn length forces the chain to spread across  multiple unit cells. The equilibrium configuration resembles that assumed by a lattice animal (see Fig. \ref{fig:LA}). 

\begin{figure}[!h]
\includegraphics[scale=0.7]{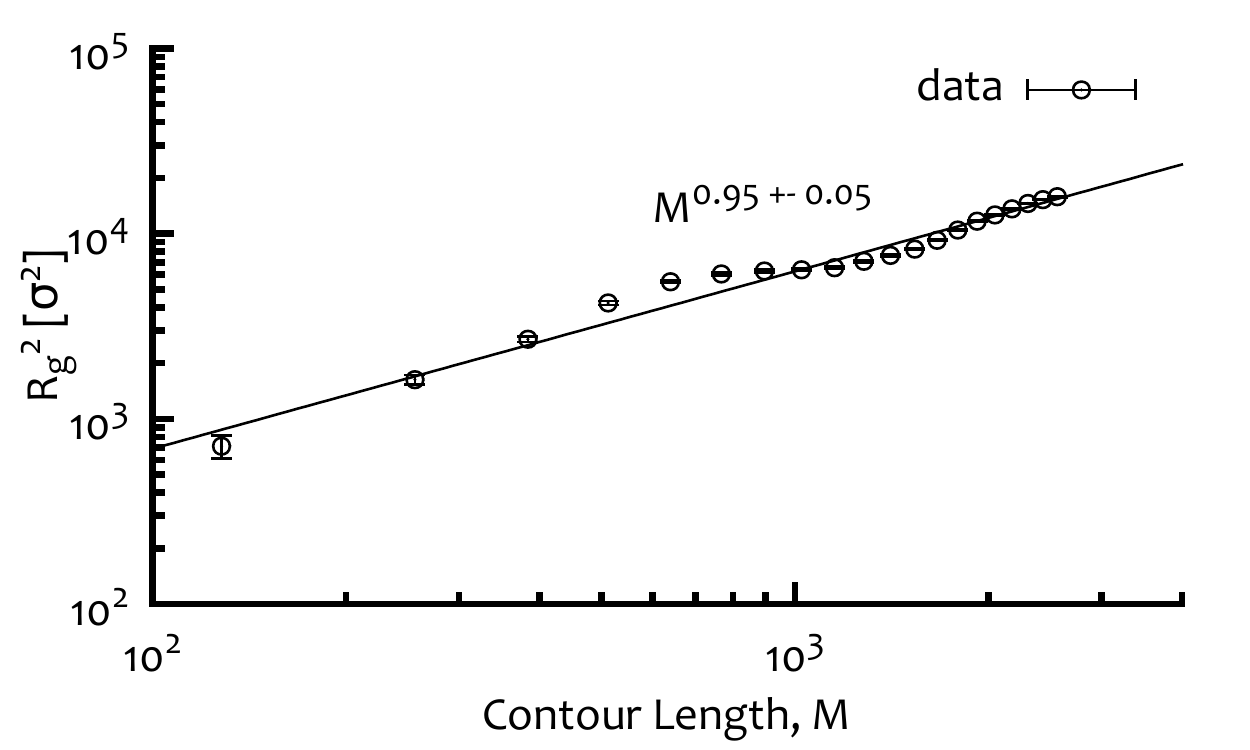}
\caption{Radius of gyration squared $R^2_g$ for different contour lengths, averaged over different starting points along the chain. The fit suggests values of the entropic exponent $\nu$ in agreement with those found by the Kinetic Monte Carlo approach described in the main text.}
\label{fig:FractDim}
\end{figure}

In order to study the scaling of the gyration radius of the polymer we compute $R^2_g$ of portions (much longer than the Kuhn length) of the chain. Our findings are shown in Fig. \ref{fig:FractDim}. We report a scaling law $R^2_g \sim M^{0.95 \pm 0.05}$ that is in agreement with the values of $\nu$ found in the main text. 

\end{document}